\PassOptionsToPackage{dvipsnames,svgnames,x11names}{xcolor}
\documentclass[sigconf,noacm]{acmart}

\usepackage{amsmath,amsfonts}
\usepackage{graphicx}
\usepackage{textcomp}
\usepackage{pifont}
\usepackage{multirow}

\usepackage{url}

\usepackage{tikz}
\usepackage{pgfplots}
\usepackage{subcaption}
\usepackage{algorithm}
\usepackage{algpseudocode}
\usepackage{verbatim}

\newcommand{\cmark}{\textcolor{ForestGreen}{\ding{51}}}%
\newcommand{\xmark}{\textcolor{red}{\ding{55}}}%

\AtBeginDocument{%
  }

\setcopyright{none}
\renewcommand{\footnotetextcopyrightpermission}[1]{}
\acmConference{}{}{}{}
\renewcommand{\acmConference}[4]{}

\begin{document}

\title{ELiTeFormer: An Efficient Transformer for FPGAs}

\author{Victor Agostinelli}\authornote{Currently at Meta.}
\affiliation{%
  \institution{Oregon State University}
  \city{Corvallis}
  \state{Oregon}
  \country{USA}
}
\affiliation{%
  \institution{Pacific Northwest National Laboratory}
  \city{Richland}
  \state{Washington}
  \country{USA}
}
\email{agostiniv@oregonstate.edu}

\author{Nicolas Bohm Agostini}
\affiliation{%
  \institution{Pacific Northwest National Laboratory}
  \city{Richland}
  \state{Washington}
  \country{USA}
}
\email{nicolas.agostini@pnnl.gov}

\author{Antonino Tumeo}
\affiliation{%
  \institution{Pacific Northwest National Laboratory}
  \city{Richland}
  \state{Washington}
  \country{USA}
}
\email{antonino.tumeo@pnnl.gov}


\begin{abstract}
Transformer blocks are prevalent in large language model (LLM) but present deployment challenges due to their challenging computational and memory demands.
While prior work has typically optimized attention mechanisms or feed-forward networks (FFNs) separately, few hardware (HW) architecture have jointly addressed both components with co-designed hardware acceleration.
We present ELiTeFormer (\textit{E}fficient \textit{Li}near \textit{Te}rnary Trans\textit{former}), the first Transformer model architecture that unifies hybrid linear attention with ultra-low-precision (ternary) linear projections, specifically co-designed for field-programmable gate array (FPGA) deployment.
ELiTeFormer achieves $10\times$ model weight compression and $12.8\times$ key-value (KV) cache compression compared to LLaMA 3, while maintaining competitive accuracy (31.9\% on the MMLU benchmark, within 3.0\% of BitNet b1.58).
Our key architectural contribution is a novel processing element (PE) micro-architecture that eliminates all multiplications in ternary linear projections through bitmasking operations, significantly reducing resource utilization by completely avoiding dedicated digital signal processing (DSP) blocks.
We simulate, synthesize, and deploy ELiTeFormer targeting a Xilinx VCK5000 Versal board using high-level synthesis (HLS) flows.
Block-level simulations show $9.6\times$ speedup for FFN operations and $4.4\times$ speedup for attention compared to standard implementations.
End-to-end deployment achieves up to $3.9\times$ lower latency and $3.2\times$ better energy efficiency than LLaMA 3 on an NVIDIA A100 graphics processing unit (GPU) at long context lengths.
This represents the first FPGA realization combining linear attention with ternary quantization, demonstrating the viability of algorithm-architecture co-design for next-generation LLM acceleration.
\end{abstract}

\maketitle


\section{Introduction}
Transformer-based large language models (LLMs) have seen explosive growth in both industry and academia, with rapid progress in model capabilities and widespread adoption across domains. LLM use spans a diverse range of applications, including question/answering and general conversational artificial intelligence (AI) \cite{openai_2024_gpt4}, assisting in image generation \cite{liu_2024_llm_diffusion}, code generation and analysis \cite{jiang_2024_survey_codegen}, basic sentiment analysis \cite{zhang_2024_sentiment}, and more. As these models scale in size and complexity, delivering efficient compute and memory solutions has become a significant challenge. This is particularly true in scenarios demanding low latency and energy efficiency, such as real-time inference and edge deployment. Even in datacenter settings, reducing the cost and energy footprint of inference at scale remains critical.

On the algorithmic side, researchers have proposed several alternatives to address inefficiencies. These efforts typically focus on optimizing individual components such as attention or feed-forward network (FFN) blocks. Efficient attention mechanisms aim to subquadratic scaling with context length, including sparse attention~\cite{child_sparsetransformer_2019, tay_sparse_2020, bigbird2020} and linear attention~\cite{kathrapalous_linear_2020, agostinelli_improving_2023, yang_gla_2024, agostinelli_leapformer_2024}. FFN efficiency has been pursued via reduced-precision arithmetic~\cite{dettmers_2022_int8, wang_bitnet1b_2023, ma_bitnet158b_2024} and fine-grained mixture-of-experts approaches~\cite{chen_2022_moe}. However, these optimizations are often explored in isolation.  To the best of our knowledge, no prior work has jointly optimized both the attention and FFN blocks with co-designed hardware architectures tailored to exploit both algorithmic innovations simultaneously. 


Recent work explored hardware acceleration for Transformer models and their lightweight variants in custom silicon or with reconfigurable devices \cite{Ham_a3_2020, wang_spatten_2021, fan_butterfly_2022, zeng_flightllm_2024}. 
Field-programmable gate arrays (FPGAs) provide the ability to implement highly customized and performant accelerators for these Transformer-based models. Application-specific integrated circuits (ASICs) generally achieve even higher performance and superior energy efficiency. However, deep learning models evolve quickly, often rendering fixed-function accelerators obsolete. The reconfigurability of FPGAs provides the flexibility needed to track these rapid shifts, rendering them attractive despite their performance gap relative to ASICs.

Prior work \cite{zeng_flightllm_2024, chen_understanding_2024} has shown that, especially for generative inference, FPGAs can outperform GPUs by exploiting fine-grained pipelining and dataflow parallelism. However, these prior FPGA-based approaches inherit inefficiencies from general-purpose Transformer designs. Notably, they attempt to support both prefilling and generative stages on a unified architecture, which leads to increased complexity and poor key-value (KV) cache management. Furthermore, they accelerate only standard attention mechanisms without leveraging algorithmic innovations in efficient attention or ultra-low-precision FFNs. These architectural and algorithmic mismatches hinder the ability to fully realize the performance and energy benefits of FPGA-based deployment.

To address these limitations, we propose a hardware-software (HW/SW) co-design approach centered around a custom, Transformer-based model optimized for both algorithmic and hardware efficiency. We introduce \textit{ELiTeFormer}—the \textit{E}fficient \textit{Li}near \textit{Te}rnary Trans\textit{former}—a novel architecture designed to overcome the bottlenecks observed in existing FPGA-targeted LLM deployments. For the first time, ELiTeFormer jointly optimizes the attention and FFN blocks by combining recent advances in efficient attention and extremely low-precision weights, and co-designs specialized hardware to exploit these algorithmic innovations. It employs a hybrid-linear attention mechanism with a sliding window, trained via \texttt{LoLCATs} \cite{zhang_2024_lolcats} for efficient long-context modeling via distillation. Simultaneously, it leverages BitNet b1.58-style ternary-quantized linear projections \cite{ma_bitnet158b_2024} in both the attention and FFN components. These low-precision operations are well-suited for FPGAs, which can natively support arbitrary bit-width computation and flexible datapaths. 
To exploit this, we develop a custom processing element (PE) micro-architecture that eliminates multipliers entirely, replacing them with efficient bitmasking operations. The design is synthesized using high-level synthesis (HLS) flows \cite{ferrandi_2021_bambu,amd_2025_vitishls}, simulated for performance analysis, and deployed on Xilinx FPGAs.
This work makes the following key contributions:
\begin{itemize}
    \item We propose \textbf{ELiTeFormer}, to the best of our knowledge, the first Transformer-based model that jointly combines hybrid-linear attention and ultra-low-precision linear projections. This enables 10$\times$ model weight compression and 12.8$\times$ or more KV cache compression in long-context scenarios compared to LLaMA 3 \cite{grattafiori2024llama3herdmodels}, a state-of-the-art open-weight LLM. We accomplish this while incurring only a 3.0\% accuracy drop on the MMLU benchmark \cite{eval-harness}
    \item We present \textbf{ELiTeFormer PE} (ELTF PE), a novel PE micro-architecture that accelerates multiply-and-accumulate (MAC) operations for BitNet b1.58-styled linear projections. ELTF PE avoids ternary multiplication via bit-masking, eliminating the usage of dedicated FPGA DSP blocks for linear projections, and is the first ternary PE applied towards Transformer workloads.
    \item We synthesize, simulate, and deploy an accelerator platform specialized for ELiTeFormer using HLS flows \cite{ferrandi_2021_bambu,amd_2025_vitishls} on Xilinx FPGA-based data-center accelerator boards. To the best of our knowledge, this is the first custom hardware design and FPGA realization that includes \textit{both} an efficient linear-attention and BitNet b1.58-styled linear projections.
    \item We demonstrate that ELiTeFormer achieves significant latency improvements in simulation and deployment. We achieve a $9.6\times$ speedup on the linear projections and a $4.4\times$ speedup on the attention block when compared to quadratic references in simulation.
    On hardware, when deploying a scaled-down ELiTeFormer on a Xilinx VCK5000 board, we achieve up to a $3.9\times$ latency speedup with long context lengths and $3.2\times$ better average energy efficiency than a similarly scaled version of a LLaMA 3 LLM running on an NVIDIA A100 GPU.
\end{itemize}

\section{Background and Motivation}
In this section, we review the Transformer model design \cite{vaswani_attention_2017}, including a brief overview of the attention mechanism and key concepts in efficient, alternative operators. We also discuss a major challenge of accelerating attention on FPGAs: unbounded generation.

\subsection{Transformer Basics}
Transformers blocks are formed, fundamentally, by symmetrical layers containing attention blocks, an FFN, and normalization blocks. We focus on decoder-only implementations \cite{radford2018improving} in this example. Equation~\ref{eq:transformer} describes the output $y$ as a function of input $x$ and a single layer of the original Transformer architecture in a more formal manner, where the self-attention block is denoted as $a$, the FFN block is denoted as $FFN$, and wrapped by normalization blocks (typically LayerNorm) denoted as $LN$. 

\begin{equation}
    \begin{split}
    \bar{x}(x) = LN(x + a(x)) \\
    y(x) = LN(\bar{x} + FFN(\bar{x}))
    \end{split}
    \label{eq:transformer}
\end{equation}

The attention and FFN blocks are computationally demanding and are commonly the focus of efficiency optimizations. Decoder-only Transformers typically have two phases during inference: prefilling and decoding (hereafter called generative inference or generation). Prefilling takes a series of contextual vectors, analogous to a prompt for an LLM, and generates embedded or hidden representations that are saved in the KV cache. This context is then used for attention during generative inference, where the model generates a single token at each time-step and adds that token's hidden representations to the KV cache as context for the next time-step.

\subsection{Attention and Efficient Variants}
Here, we cover the basics of classical attention as introduced by Vaswani et. al \cite{vaswani_attention_2017} in addition to a range of efficient variants for this operator, including sparse attention \cite{child_sparsetransformer_2019} and linear attention \cite{kathrapalous_linear_2020}.

\subsubsection{Classical Attention}
\label{sec:attn}
The attention mechanism, as originally introduced in Transformers \cite{vaswani_attention_2017}
can be expressed in a simplified form for self-attention as seen in Equation~\ref{eq:attn} on a per-attention head and time-step basis (before concatenation). In this example, we define $q_{h, t} \in \mathbb{R}^{1 \times d_k}$, $K_{h} \in \mathbb{R}^{t \times d_k}$, and $V_{h} \in \mathbb{R}^{t \times d_v}$. The head dimensionality is referred to as $d_k$ and $d_v$ in this example and is called $d$ when unified for $q_{h, t}$, $K_{h, t}$, and $V_{h, t}$ for the sake of simplicity. $t$ refers to a given time-step during generation and also corresponds to the sequence length of $K_{h, t}$ and $V_{h, t}$, which contain cached tokens from previous time-steps up to $t$. $h$ corresponds to some $h \in H$ where $H$ is the set of available attention heads.

\begin{equation}
    a_{h, t} = softmax\left( \frac{q_{h, t} K^T_{h, t}}{\sqrt{d}}\right) V_{h, t}
    \label{eq:attn}
\end{equation}


Classical attention has quadratic $O(n^2)$ cost in both time and memory footprint for long sequences, driven by the $q_{h, t}K^T_{h, t}$ interaction. This has led to strong interest in more efficient alternatives.

\subsubsection{Sparse Attention}
Sparse attention leverages sparsity in the attention activations to reduce the computational complexity of attention operations down to a constant cost. Popular sparse attention variations take many forms \cite{child_sparsetransformer_2019, bigbird2020, beltagy_longformer_2020},
most approaches rely on fixed, structured sparsity patterns that map well to general-purpose hardware and reduce memory traffic.
More dynamic sparsity patterns are possible, but usually require custom hardware to materialize their theoretical efficiency gains.



\subsubsection{Linear Attention}
Another efficient attention alternative, linear attention \cite{kathrapalous_linear_2020} attempts to reformulate the enforced order of computation in Equation~\ref{eq:attn} with a lossy approximation by replacing the $exp$ in the $softmax(...)$ operator with some similarity-measuring function or kernel that is separable.
After this substitution, the attention operator's calculation can be reordered to form a linear operator with respect to the length of the sequence in terms of runtime and memory footprint. This can be observed in Equation~\ref{eq:lin_attn}, where $k_{h, t}$ represents a vector at row $t$ in $K_{h, t}$ and $\Tilde{x}$ represents an approximation of vector or matrix $x$. We drop $\sqrt{d}$ scaling for simplicity.

\begin{equation}
     a_{h, t} \approx \Tilde{a}_{h, t} = \frac{\Tilde{q}_{h, t}(\Tilde{K}^T_{h, t}V_{h, t})}{\Tilde{q}_{h, t}\sum^t_{j=0}\Tilde{k}^T_{h, j}}
    \label{eq:lin_attn}
\end{equation}

During casual training and generation, Equation~\ref{eq:lin_attn} can be modeled as a recurrent function with a constant-sized state.
Hybrid attention variants featuring linear attention alongside sparse attention \cite{zhu_long_short_2021, arora_simple_2024, dong_less_2024} have gained popularity recently and represent promising prospects in terms of model quality.

\subsection{Low-Precision FFNs}
FFNs are a common component in many neural network architectures, and their optimization has been widely studied. In particular, low-precision FFNs \cite{wang_bitnet1b_2023, ma_bitnet158b_2024, dettmers_4bit_2023} offer a promising and efficient alternative. These variants reduce memory footprints through quantized representations. When low-precision computation is supported by hardware, these compression schemes can provide significant theoretical speedups. However, these gains are rarely realized on general-purpose hardware.

One recent implementation, BitNet b1.58 \cite{ma_bitnet158b_2024}, proposes replacing the weights of large scale models with purely ternary representations 
(i.e., discretized to $[-1, 0, 1]$). While this extreme quantization results in some model quality loss, 
models at scale are often able to overcome even significant approximation error due to redundant logical pathways during inference (similar to the intuition behind why model pruning works well \cite{sun_2024_pruning}) especially with additional training. 


\subsection{A Challenge: Unbounded Generation and FPGAs}
FPGA-based Transformer accelerators often target vision tasks \cite{rahman_efficiency_2016, wang_via_2022}, partly due to their popularity and partly to avoid a key engineering challenge: static compilation on FPGAs complicates support for dynamic workloads such as unbounded generation with a growing context window \cite{zeng_flightllm_2024}.
One common approach for general acceleration is to tile input data using a host processor during the prefilling stage \cite{chen_understanding_2024, li_ftrans_2020, khan_npe_2021}. However, this still requires provisioning large buffers to handle a range of possible key-value matrix sizes during attention, which can be costly \cite{chen_understanding_2024, li_ftrans_2020}. Another explored solution is to design a custom instruction set architecture (ISA) that supports different datapath configurations. While flexible, this approach can introduce significant overhead. For example, FlightLLM \cite{zeng_flightllm_2024} requires 3.25 GB of DRAM just to support its ISA.


In general, existing techniques for prefilling in Transformer-based LLMs on FPGAs adequately address concerns with high utilization and middling waste. 
Other works even propose employing heterogeneous systems for prefilling \cite{chen_understanding_2024}, focusing on aligning architectural strengths of different accelerators with different workload phases.
Unfortunately, existing designs that attempt to unify prefilling and generation within the same hardware solution are necessarily inefficient during generation, as generation imposes additional overhead to support different runtime configurations.
Furthermore, adapting these approaches is non-trivial, as many are handcrafted and tightly coupled to low-level implementations, making them difficult to retarget or generalize with HLS flows that leverage high-level algorithmic descriptions.

Ultimately, we argue that current efforts to accelerate attention during generation rely on algorithms that are mismatched to FPGA architectures, overlooking alternatives that align more naturally with the strengths of fine-grained reconfigurable hardware.  
This understanding should underscore the HW/SW co-design philosophy and efforts advocated for in recent work~\cite{zeng_flightllm_2024, chen_understanding_2024}.


\begin{figure*}
    \centering
    \includegraphics[width=1\linewidth]{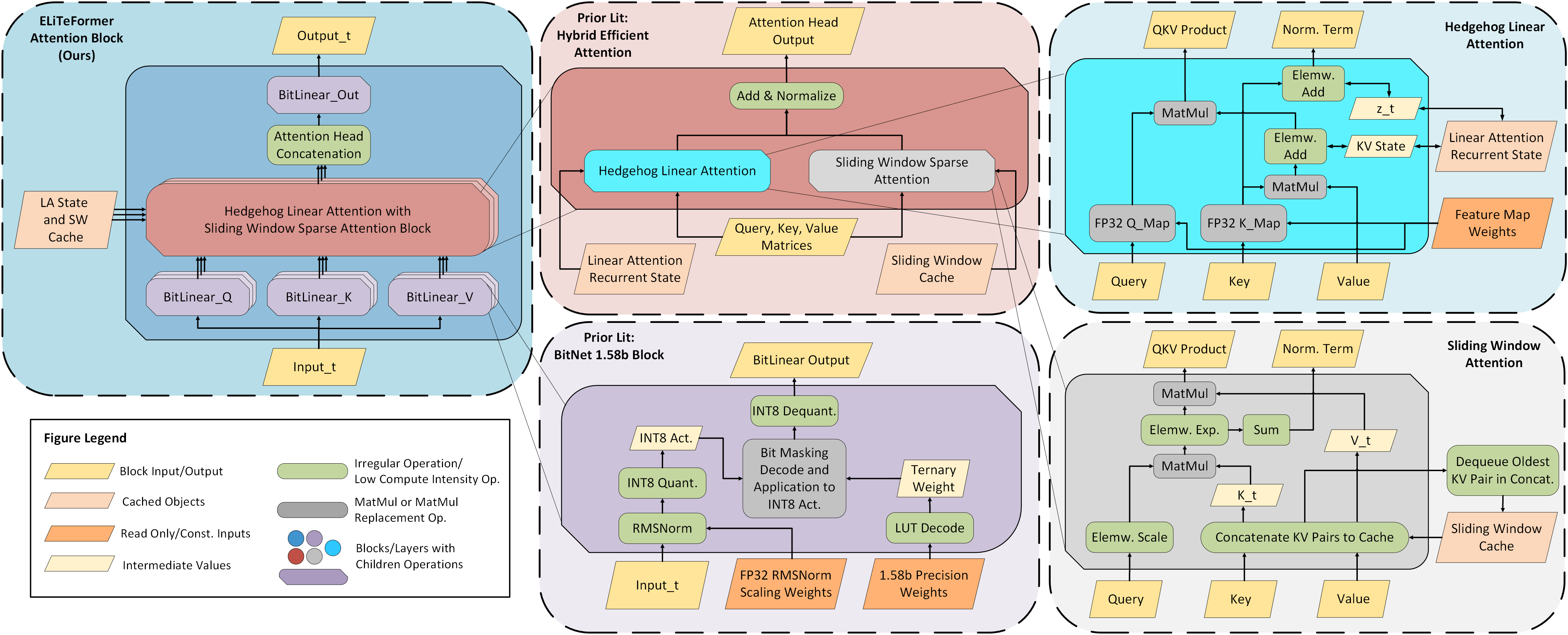}
    \caption{Depiction of ELiTeFormer's attention block during generative time-steps. 
    Hedgehog Linear Attention \cite{zhang_hedgehog_2024} + Sliding Window Sparse Attention \cite{child_sparsetransformer_2019} are executed in parallel for each attention head.}
    \label{fig:ELiTeFormer_attn}
\end{figure*}

\section{ELiTeFormer Model and FPGA  Implementation}

As mentioned previously, accelerating efficient Transformer variants in hardware is not new and is sometimes considered critical for the viability of a proposed solution. For example, fairly dynamic attention sparsity can be better exploited in ASIC and FPGA-based solutions than when an architecture is deployed on general-purpose hardware \cite{fan_butterfly_2022, zeng_flightllm_2024, shen_salo_2022}. However, prior works have not necessarily modified algorithms and/or model architectures enough to guarantee suitability for a given custom hardware platform. Moreover, such works often attempt to leverage FPGAs for prefilling when GPUs are already well-suited for the task. As such, we focus only on generation moving forward and assume ELiTeFormer is deployed in a heterogeneous environment that features separate acceleration platforms for prefilling and generation.


\subsection{An Algorithmic Fit for FPGAs}
When it comes to FPGAs, two factors are critical for a given Transformer algorithm variant to be easily adaptable and viable. First, a given variant should be computationally \textit{efficient} in that it should avoid abusing runtime and memory footprints. At large deployment scales, more efficient attention and FFN variants can iterate more quickly, process multiple queries simultaneously via large batches, and potentially save on energy costs. Alternatively, at smaller scales or on the edge, hard runtime and memory constraints may be required to be met. Additionally, a given variant should be near or completely \textit{static} in its memory footprint. This eliminates any possible waste from length-adaptive overhead required to support unbounded generation. 

From this perspective, linear attention and specific sparse/linear hybrid architectures represent interesting prospects. As referenced in Section~\ref{sec:attn}, the computational profile of linear attention during generation is determined purely by model hyperparameters. It is thus calculable \textit{a priori} for static compilation. Some sparse/linear hybrid attention variants are similarly determined \cite{arora_simple_2024, dong_less_2024}. These architectures are useful for their \textit{efficiency} and are demonstrably \textit{static}.
As such, we believe that they should be more widely used on FPGAs for generative workloads.

Unfortunately, after replacing an attention operator with an efficient alternative, further efficiency gains are usually limited without also optimizing or substituting the basic Transformer FFN block.
Thus, we observe that optimizations here are somewhat symbiotic. Efficiency-focused substitutions in one block encourage efficiency-focused substitutions in the other, resulting in additional speed-ups and reduced memory footprints. For this reason, we also propose applying BitNet b1.58 \cite{ma_bitnet158b_2024} to the FFN block, representing the first time a Transformer-based model has featured both an efficient attention mechanism and a ternary FFN and, to the best of our knowledge, the first time that linear attention or BitNet b1.58 has been synthesized and deployed on an FPGA.

\subsection{Pushing the Pareto Frontier of Efficient Models at Scale}
\label{sec:ELiTeFormer_pareto}

When combining attention and FFN architectural optimizations, we can push the Pareto frontier of efficiency. 
We propose a novel, ultra-efficient generative model architecture, which we call \textit{ELiTeFormer}. Our Transformer-based model combines the data-center scalability and wasteless structure of linear attention during generative inference with the vastly increased speed and reduced model memory footprint of low-precision FFNs.
For the baseline model, we employ LLaMA 3 8B for several key reasons: it is a state-of-the-art open-source model architecture, widely adopted across industrial and academic settings, and there is a BitNet b1.58 checkpoint\footnote{The BitNet b1.58 checkpoint we utilize was pre-trained on 100B tokens. Based on the reported training efficiency of LLaMA 3 8B, this requires roughly 8,700 GPU hours to produce the parameter-efficient model.} based on LLaMa 3 8B that we adapt for ELiTeFormer.

An ELiTeFormer attention block is depicted in Figure~\ref{fig:ELiTeFormer_attn}, which outlines a hybrid linear attention and sliding window architecture that has demonstrated success in prior work \cite{zhang_2024_lolcats, arora_simple_2024}, but with the typical attention head projections replaced with low-precision alternatives based on BitNet b1.58 \cite{ma_bitnet158b_2024} and a classical sliding window implementation versus a ``terraced'' alternative \cite{zhang_2024_lolcats}. As part of our hybrid attention, additional learned feature maps are included \cite{zhang_hedgehog_2024} in full precision to maintain model stability, adding approximately 1\% additional parameters to the base parameters for ELiTeFormer. Low-rank adapters (LoRAs) \cite{hu_2022_lora} are also typically included in these types of adapted attention blocks. We choose not to include adapters in the case of ELiTeFormer, as we found them ineffective (elaborated upon in Section~\ref{sec:ELiTeFormer_attn_distill}). This enables ELiTeFormer to maintain a static inference profile for all context lengths, a significant improvement in terms of memory footprint, particularly at long contexts.

The ELiTeFormer FFN block and attention linear projections (see BitLinear blocks in Figure~\ref{fig:ELiTeFormer_attn}) are taken directly from BitNet b1.58, employing the same low-precision structure and ternary quantization scheme. This is particularly interesting for FPGA deployment, as arithmetic operations featuring extremely low-precision values are usually poorly supported on traditional architectures. Moreover, ternary weights offer the opportunity to completely eliminate multiplication operations and only accumulate during linear projections, which is almost impossible to exploit in general-purpose hardware. 

\begin{figure}
    \centering
    \includegraphics[width=\linewidth]{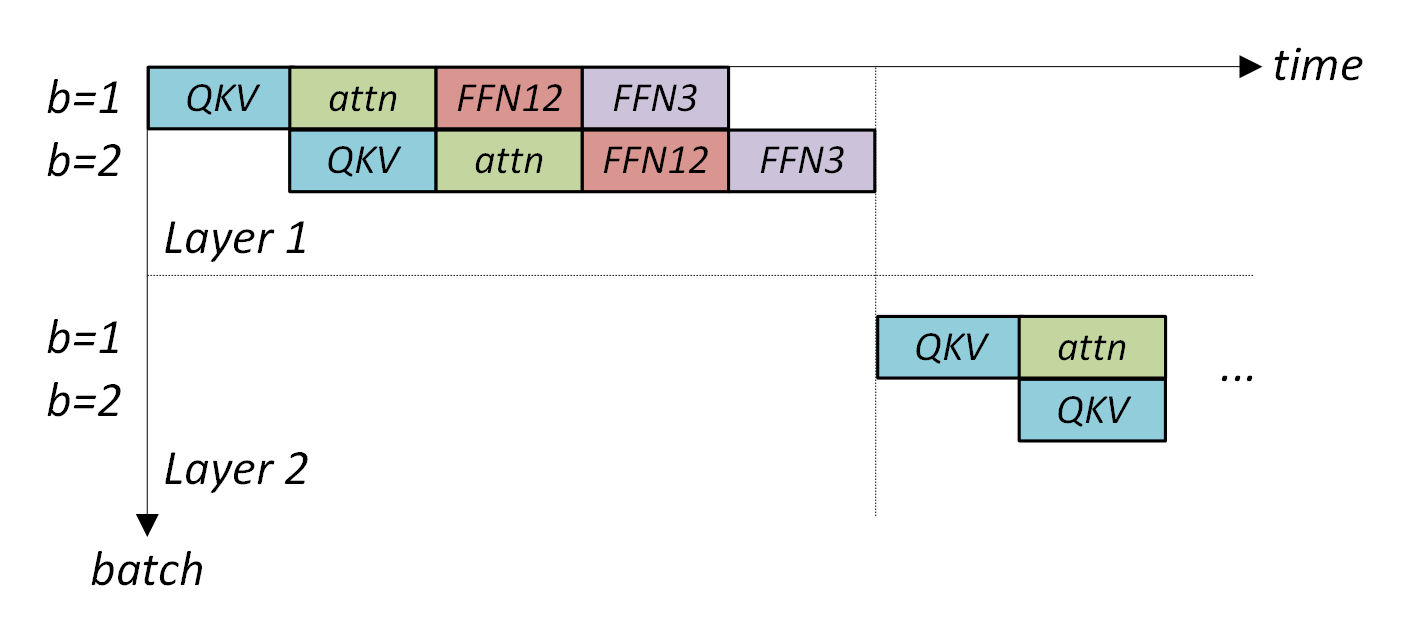}
    \caption{An example illustration of the timing diagram for the ELiTeFormer top-level dataflow, pipelining the generative inference stages across two batches. Stages are load balanced to have similar latency.}
    \label{fig:dataflow_illustration}
\end{figure}

\subsection{Defining ELiTeFormer Inference on FPGAs}
\label{sec:pipeline}
To design ELiTeFormer for efficient inference on FPGAs, we draw inspiration from the analytical performance models provided by Chen et al. \cite{chen_understanding_2024}, which we adapt for our customized model architecture and deployment environment.
The top-level design is pipelined per-layer into the following discrete stages, where each stage operates on a batch of samples. We define query/key/value projections with positional embeddings and the half-precision per-head linear projections required for linear attention as $QKV$. Jointly, we define the operations implementing linear attention, sliding window attention, and the attention output projection as $attn$.  As is necessary for LLaMA 3-styled models, we also define the first parallel halves (colloquially called up and gate projections) of the FFN as $FFN_{1, 2}$, and the second half of the FFN as $FFN_3$.

This pipeline is depicted in Figure~\ref{fig:dataflow_illustration}. Notably, the proposed dataflow architecture differs slightly from the one proposed by Chen et al. \cite{chen_understanding_2024}, not just in how stages are constructed, but also by another critical factor: all stages have synchronization boundaries, and computation cannot be overlapped.
In such settings, commonly used normalization layers introduce synchronization points because the full token output is required to compute statistical properties, such as the mean or variance. Since the BitNet b1.58 projections \cite{ma_bitnet158b_2024} are prepended with a RMSNorm block in order to improve model accuracy, full token-level computation is required before a given stage featuring such a projection can begin. 

\subsection{FPGA Design and Ternary Precision PE $\mu$-arch}
\label{sec:implementation}

ELiTeFormer and its accelerator platform was carefully co-designed to target the Vitis HLS and Vivado design flow. Here, we review the design choices employed in the proposed ELiTeFormer's FPGA implementation, with a focus on the memory interface specification and the ELiTeFormer's low-precision PE micro-architecture, which we refer to as \textit{ELTF PEs} for BitNet b1.58-styled linear projections.

\begin{figure*}
    \centering
    \includegraphics[width=1\linewidth]{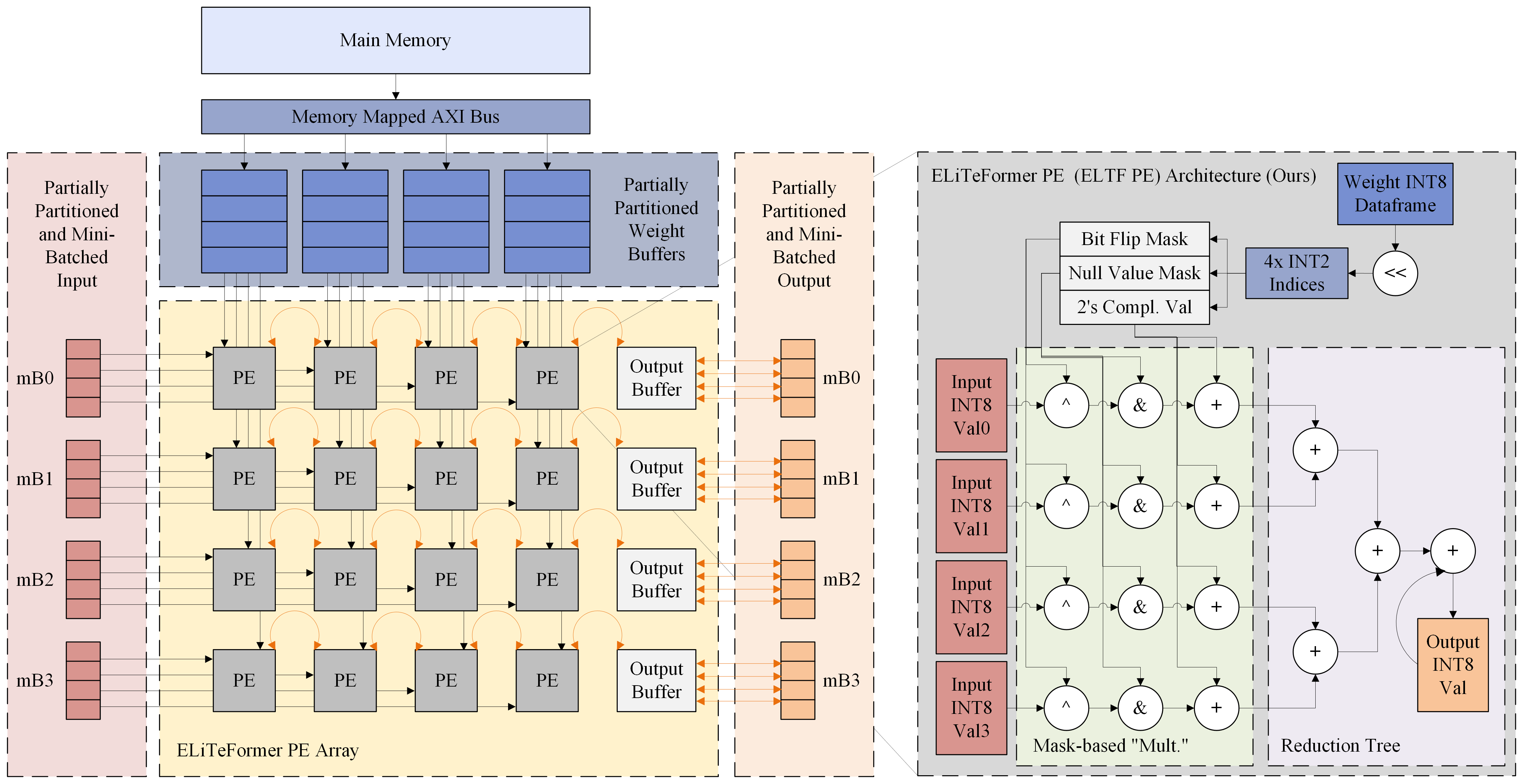}
    \caption{An illustration of the proposed ELiTeFormer linear projection architecture (left) and PE-specific architecture (right) in hardware. On the left, the input, output, and weight matrices are partially partitioned to enable parallel accesses. The input and output are also split into mini-batches $mB$ and enumerated as $mBx$, with the example in this figure setting $mB = 4$. Data transfers related to the output are color-coded on the left, represented by orange, indicating the wide availability of the output buffer to the PEs. On the right, a bitwise mask-based replacement for multiplication is employed for efficiency, where unpacked 2-bit weights index the masks. Each weight dataframe is compressed column-wise, so a reduction tree is employed to modify the associated output value.}
    \label{fig:bitnet_pe}
\end{figure*}

\subsubsection{Memory Interface Specification and General Notes}
Regarding our memory interface, we employed typical memory-mapped AXI interfaces for the model weights. This type of interface supports bus widths of up to 512 bits and can burst-read up to 4 kB at a time. Most weights are packed into 8-bit dataframes, supporting 4 weights per dataframe (8x compression), which could be further improved with more complex look-up table (LUT) decoding, such as a 5/3 scheme using 30/32 bits in a 32-bit dataframe. Consequently, we read 256 weights in a single read before bursting. 
Some smaller weight matrices (e.g., normalization weight vectors) that need to remain in 32-bit floating point are loaded along similarly sized memory buses.

The ELiTeFormer attention module accesses both its linear-attention state and sliding-window cache through similar interfaces. Although full parallelization across all attention heads is possible, it depends on platform resources. After applying our optimizations, the operator runs efficiently even when heads are processed sequentially. Achieving full head-level parallelism requires adequate memory banking and compute capacity. Moreover, the $attn$ stage already consumes a large share of system resources (elaborated later in Section~\ref{sec:deployment}). For this reason, aggressive parallelization can be impractical. In our later experiments, we instead partially parallelize the query heads in accordance with a single key-value head in grouped-query attention (GQA) (slight variant of attention in LLaMa~3 models where $H_q > H_{k,v}$). For example, when four query heads map to a single key-value attention head, we compute all four query attention heads in parallel. This approach increases compute usage but does not add pressure to the platform’s off-device memory interface.

In general, all computational blocks followed a dataflow structure, leveraging co-execution of different stages when multiple iterations are in-flight and dependencies are met. Every major block follows the same pattern: (1) load relevant data from off-chip memory in 512-bit frames, (2) decode the 512-bit frames into their intended datatypes, (3) engage in potentially multi-stage computation, and sometimes (4) encode and write back to off-chip memory in 512-bit frames. The final write-back stage only occurs for linear attention recurrent state modifications and in the top-level dataflow. In contrast, the output of every other stage can be stored in (on-chip) block RAM (BRAM) to avoid off-chip accesses.




\subsubsection{ELiTeFormer Linear Projections and PE $\mu$-arch}
\label{sec:bitnet_proj}

Our proposed architecture for ELiTeFormer linear projections can be observed in Figure~\ref{fig:bitnet_pe}, where the left half of the figure articulates the roughly systolic structure of the PE array and its memory interfaces, and the right half defines the specialized ELiTeFormer PE micro-architecture, which is a novel PE micro-architecture. 
A mini-batch dimension, which we denote as $mB$, improves parallelism in the input and output matrices of these linear projections. These projections are stored in BRAM blocks of size $mB \times d_{model}$ each and are then partitioned to enable parallel access. 
This additional mini-batch dimension significantly improves weight reuse at the expense of increased hardware resources, thereby proportionally reducing latency and increasing throughput in multi-batch inference scenarios. In reality, the increased resource cost is also not particularly constraining, as this optimization is only really applicable to non-attention operations (since attention caches scale with $mB$ whereas linear projections do not). This will be elaborated upon in Section~\ref{sec:deployment}.


The right half of Figure~\ref{fig:bitnet_pe} elaborates on the novel micro-architecture of the proposed ELTF PEs that efficiently accelerate BitNet b1.58-styled linear projections.
The ELTF PEs first read the broadcasted input values for the implemented compression scheme, which represent 4 input values in this example\footnote{We employ a simplified 8x compression scheme. A more aggressive scheme (up to 10x theoretical compression) would require more input values to be read and a larger weight dataframe.}. These input values are held stationary in the PEs over time until all relevant weights have been used. After the input values are loaded (in parallel to the corresponding output value being loaded), the 8-bit weight dataframe is loaded from the partitioned weight buffers and bit-shifted into four indices, which are used to select specific masks for our proposed mask-based multiplication emulation. Since we are employing ternary weights (i.e., in the range [-1, 0, 1]), the ELTF PEs only need to nullify a value, flip its sign, or allow it to pass through for accumulation to correctly emulate multiplication, which is accomplished via the application of the indexed masks in corresponding bit-wise operators. Finally, the results of these bitwise operations are accumulated via a reduction tree, combined with the output value, and then written back to the appropriately partitioned output buffer, which eventually writes to BRAM.

\subsubsection{When Does ELiTeFormer Perform Best?}
The unique inference profile of ELiTeFormer naturally lends itself to environments that require high energy efficiency and low latency, especially in long-context scenarios. Its extremely compressed memory footprint for both the model weights and the attention cache renders it ideal for constrained systems. However, ELiTeFormer is not limited to those environments.

At long contexts, cutting-edge LLM models still bottleneck their throughput on general-purpose hardware due to large attention caches taking up significant space in DRAM. This can limit the size of the inference batch on the hardware in question, lowering throughput even if the hardware is capable of parallelizing batches of that size on-chip. Using LLaMA 3 8B as an example, with a batch size of 128 at 4096 tokens of context, the attention cache takes up nearly 64 GB of DRAM. That attention cache is a little more than four times the size of the model itself. ELiTeFormer algorithmic changes to attention not only reduce its latency, but also drastically reduce the size of the attention cache. Using the same example inference parameters, ELiTeFormer's attention cache would require only 5 GB of DRAM, representing a $12.8\times$ improvement.

Naturally, the above observation presents another opportunity for ELiTeFormer: high-throughput inference at extremely large batch sizes. Whereas LLaMA 3 8B on an NVIDIA A100 80 GB would likely be limited to batch sizes of around 128, ELiTeFormer would theoretically be capable of supporting batches of over 16k, assuming similarly sized DRAM. This operation mode is possible due to algorithmic changes to the attention block, so any model leveraging hybrid attention, such as ELiTeFormer,  would be capable of it. However, coupling it with ternary precision linear projections enables ELiTeFormer to achieve lower latency than those alternatives.

\section{Results}
We evaluate ELiTeFormer along two axes: model quality under aggressive efficiency constraints, and the extent to which its expected hardware advantages appear in practice. For training and evaluation, we fine-tune a BitNet b1.58 8B checkpoint \cite{ma_bitnet158b_2024}, adapting the attention-distillation workflow from \texttt{LoLCATs} \cite{zhang_2024_lolcats} in PyTorch to produce an ELiTeFormer-8B model.
During training and evaluation, we used a single NVIDIA A100 GPU per run. Training employed a standard 50k slice of samples from the \texttt{LoLCATs} version of the Alpaca dataset\footnote{\url{https://huggingface.co/datasets/yahma/alpaca-cleaned}} \cite{alpaca}. Models were trained with batch sizes of 8 and sequence lengths of 1024 tokens. Learning rates started at 2e-5 before decaying via an inverse square scheduler to a minimum of 1e-7. 4000 weight updates of training (around 2.5 epochs) were required across the 50k slice of samples before convergence was observed.

For evaluation in hardware,  including simulations, we provide results for comparisons between baseline Transformer computational blocks and our proposed ELiTeFormer blocks in simulation via Bambu HLS \cite{ferrandi_2021_bambu}, an open-source HLS flow. This functions as a limited ablation study. We also provide platform-wide simulations and deploy a scaled-down version of our proposed platform to an FPGA via Xilinx's Vitis HLS and Vivado flows \cite{amd_2025_vitishls}.

\subsection{ELiTeFormer Attention Distillation and Model Quality Evaluation}
\label{sec:ELiTeFormer_attn_distill}
Training extremely low-precision neural networks, especially at scale, is a non-trivial task \cite{wang_bitnet1b_2023, ma_bitnet158b_2024}. Furthermore, it is not immediately apparent how to adapt such a low-precision model to employ an efficient attention block. Towards this end, we utilize the \texttt{LoLCATs} framework \cite{zhang_2024_lolcats} for attention distillation, incorporating some custom adjustments.

\begin{table}[t]
    \centering
    \caption{Perplexity values (lower is better) on the development set of the Alpaca dataset slice, comparing checkpoints trained and evaluated with varying sliding window sizes. For efficient notation, T64 refers to a checkpoint trained with a sliding window of 64 tokens, and E64 refers to a checkpoint being evaluated with a sliding window of 64 tokens, where the two values can differ. Best result is bolded.}
    \begin{tabular}{l|c}
         Experiment Setting & Dev. Set PPL \\
         \hline
         ELiTeFormer T256 \& E64 & 14.84 \\
         ELiTeFormer T256 \& E128 & 11.94 \\
         ELiTeFormer T256 \& E256 & \hphantom{0}9.15 \\
         \hline
         ELiTeFormer T64 \& E64 & 14.52 \\
         ELiTeFormer T64 \& E128 & 11.62 \\
         ELiTeFormer T64 \& E256 & \textbf{\hphantom{0}8.97} \\
    \end{tabular}
    
    \label{tab:sw_findings}
\end{table}

\begin{table}[t]
    \centering
        \caption{Perplexity values (lower is better) on the development set of the 50k sample slice from the cleaned Alpaca dataset comparing checkpoints trained with and without the secondary low-rank fine-tuning step. Best result is bolded.}
    \begin{tabular}{l|c}
         Epochs of Fine-tuning & Dev. Set PPL \\
         \hline
         ELiTeFormer @0.5 FT epochs & 73.80 \\
         ELiTeFormer @0.1 FT epochs & \hphantom{0}9.24 \\
         ELiTeFormer @0.0 FT epochs & \textbf{\hphantom{0}8.97} \\
    \end{tabular}
    \label{tab:lora_proof}
\end{table}

\begin{table*}
\centering
\caption{Various LLM architectures compared on common language modeling benchmarks with scores provided out of 100. All are based on the LLaMA 3 model with non-ternary model weights and activations being in bfloat16 precision. Where possible, scores have been reused from existing publications \cite{ma_bitnet158b_2024, zhang_2024_lolcats, grattafiori2024llama3herdmodels}, although some scores have been regenerated if experimental differences or unavailable data was observed. Attention cache scaling during generative inference is defined via model and inference hyperparameters: $B$ is the batch size, $N$ is the context length, $h$ is the number of attention heads, $w$ is the size of the local sliding window, and $d$ is the attention head dimensionality where usually $N >> w$ and $N >> d$.} 
\resizebox{\textwidth}{!}{%
    \begin{tabular}{l|ccc|cccccc|cc}
         Model Architecture & Efficient & Model & Cache & PiQA & ARC-e & ARC-c & HellaSw. & Wino- & MMLU & Avg. & Avg. \\
         and Quant. Method & Attention & Size & Scaling & & & (norm) & (norm) & grande & (5-shot) & & w/o \\
          & & & & & & & & & & & MMLU \\
         \hline
         Llama 3 8B & \xmark & \textcolor{red}{14.9 GB} & \textcolor{red}{$2BNhd$} & 79.9 & 80.1 & 53.3 & 79.1 & 73.1 & 66.6 & 72.0 & 73.1 \\
         Llama 3 8B Hedgehog & \cmark & \textcolor{red}{14.9 GB} & \textcolor{LimeGreen}{$Bhd^2$} & 77.4 & 71.1 & 40.6 & 66.5 & 54.3 & 24.2 & 55.7 & 62.0 \\
         Llama 3 8B Hedgeh. w/ SW & \cmark & \textcolor{red}{14.9 GB} & \textcolor{ForestGreen}{$Bh(2wd + d^2)$} & 80.9 & 81.6 & 55.1 & 79.3 & 72.7 & 50.7 & 70.0 & 73.9 \\
         \hline
         BitNet b1.58 8B & \xmark & \textcolor{ForestGreen}{\hphantom{0}1.5 GB} & \textcolor{red}{$2BNhd$} & 75.8 & 74.2 & 43.7 & 68.6 & 58.0 & 34.9 & 59.2 & 64.1 \\
         ELiTeFormer 8B (Ours) & \cmark & \textcolor{ForestGreen}{\hphantom{0}1.5 GB} & \textcolor{ForestGreen}{$Bh(2wd + d^2)$} &  72.5 & 70.4 & 37.0 & 58.0 & 54.6 & 31.9 & 54.0 & 58.5 \\
    \end{tabular}%
}
    \label{tab:model_quality}
\end{table*}

\subsubsection{Attention Distillation for ELiTeFormer}
Fundamentally, \texttt{LoLCATs} functions via two subsequent training phases. In the first phase, a given architecture has its attention block replaced with an efficient attention block and has all of its weights frozen except for those involved directly in the newly replaced block
Then, the architecture is trained block-by-block with a goal of minimizing output reconstruction error, as described by Equation~\ref{eq:attn_reconstruction}. This is done on a per-attention head basis, where $m$ is a given layer in the model, $a \in \mathbb{R}^{1 \times d}$ is the reference attention output vector on a per time-step basis from the original model functioning as a teacher, and $\Tilde{a} \in \mathbb{R}^{1 \times d}$ is the approximated output from the student model whose attention block has been replaced with an efficient alternative. The final, model-wide loss is averaged from the per-head losses \cite{zhang_2024_lolcats}.

\begin{equation}
l^{h, m}_{MSE} = \frac{1}{d} \sum_{i=1}^d(a_{i, h, m} - \Tilde{a_{i, h, m}})^2
\label{eq:attn_reconstruction}
\end{equation}

While uncompressed model architectures appeared to show no significant degradation down to a window size of 64 tokens in sliding-window attention, we found that the ELiTeFormer model quality with a sliding window of this size was too poor to be useful. This is demonstrated by the perplexity values observed in Table~\ref{tab:sw_findings}. However, increasing the size of the sliding window to 256 tokens significantly improved results. Furthermore, training with a smaller sliding window than that used during evaluation also significantly improved the resulting model, seemingly acting as a form of regularization.

For the second phase, the model being trained will typically have low-rank adapters, or LoRAs \cite{hu_2022_lora}, added to the input and output projections of the attention blocks before the model undergoes some light fine-tuning to enhance inter-layer error robustness. While this works well for most distillation flows, ELiTeFormer checkpoints were particularly sensitive to low-rank adjustments, as observed in Table~\ref{tab:lora_proof}, where further fine-tuning after attention distillation only deteriorated model quality. A range of experiments modifying hyperparameters yielded no improvements. We posit that this occurs because the model has more or less converged during attention distillation, and that low-rank adjustments to linear projections are likely too coarse-grained for such an extremely compressed model, whose weights need to be carefully tuned.

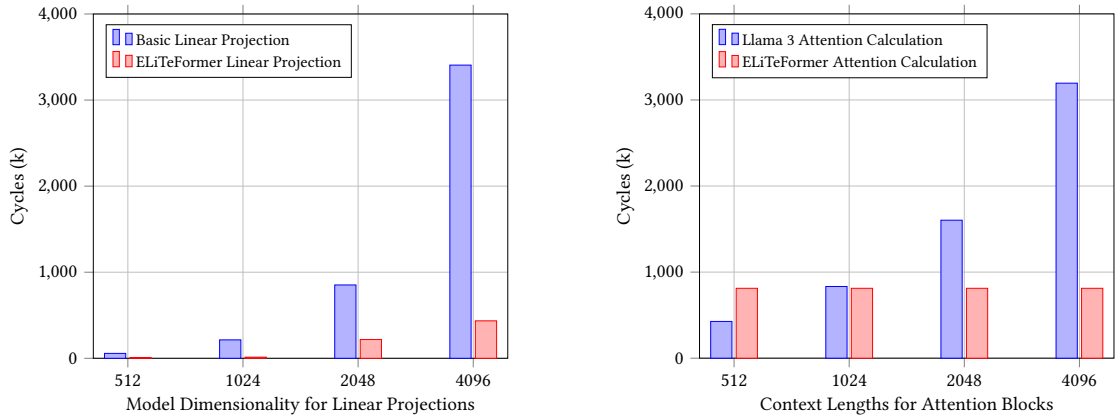
\begin{figure*}
\centering
\begin{tikzpicture}[scale=0.8]
\begin{axis}[
    ybar,
    ymax=4000,
    ymin=0,
    xlabel={Model Dimensionality for Linear Projections}, 
    ylabel={Cycles (k)}, 
    legend pos=north west, 
    tick align=outside,
    ytick pos=left,
    grid=major,
    legend cell align={left},
    legend style={nodes={scale=0.85, transform shape}},
    symbolic x coords={512,1024,2048,4096},
    xtick=data,
]
    \addplot coordinates{
        (512, 57)
        (1024, 214)
        (2048, 852)
        (4096, 3406)
    };
    \addplot coordinates{
        (512, 7)
        (1024, 13)
        (2048, 218)
        (4096, 435)
    };
    \legend{Basic Linear Projection, ELiTeFormer Linear Projection}
\end{axis}
\end{tikzpicture}
\qquad \qquad
\begin{tikzpicture}[scale=0.8]
\begin{axis}[
    ybar,
    ymax=4000,
    ymin=0,
    xlabel={Context Lengths for Attention Blocks}, 
    ylabel={Cycles (k)}, 
    grid=major,
    legend pos=north west, 
    tick align=outside,
    ytick pos=left,
    legend cell align={left},
    legend style={nodes={scale=0.85, transform shape}},
    symbolic x coords={512,1024,2048,4096},
    xtick=data,
]
    \addplot coordinates{
        (512, 427)
        (1024, 833)
        (2048, 1603)
        (4096, 3195)
    };
    \addplot coordinates{
        (512, 812)
        (1024, 812)
        (2048, 812)
        (4096, 812)
    };
    \legend{Llama 3 Attention Calculation, ELiTeFormer Attention Calculation}
\end{axis}
\end{tikzpicture}
\caption{A simulated comparison of latency in cycles (k) (lower is better) between a typical linear projection block and ELiTeFormer's low-precision alternative for a single forward pass during generative inference (left) in addition to a comparison of per-head latency in the $attn$ stage (with the output projection removed) of Llama 3 and its ELiTeFormer alternative (right).}
\label{fig:simulation_results}
\end{figure*}

\subsubsection{Evaluation of ELiTeFormer Model Quality}
Beyond development set perplexity, we evaluate the final ELiTeFormer 8B checkpoint on a range of popular language modeling benchmarks, reusing the framework provided by EleutherAI for plug-and-play LLM evaluation \cite{eval-harness}. These results can be observed in Table~\ref{tab:model_quality}, where we compare ELiTeFormer against similar model architectures. All models in Table~\ref{tab:model_quality} are derived from the Llama 3 8B checkpoint.

Generally, ELiTeFormer performs reasonably well on easier tasks in this suite of benchmarks while achieving both 10x compression like BitNet b1.58 8B \cite{ma_bitnet158b_2024} and the fast, compressed attention of Hedgehog \cite{zhang_hedgehog_2024}, utilizing a 256-token sliding window. On more challenging tasks, particularly HellaSwag and ARC-c, there is noticeable degradation compared to less efficient alternatives. For HellaSwag in particular, this may indicate a weakness in adversarial reasoning tasks. However, it is important to note that the MMLU results (where MMLU is commonly considered the most difficult of these benchmarks) were very close to those produced by ELiTeFormer's initial checkpoint, BitNet b1.58 8B, indicating that our modified distillation method is performing well.

It is worth mentioning that the necessary pretraining approaches for these models were different and favored the Hedgehog checkpoints. The explicit LLaMA 3 8B models and the efficient attention alternatives were pre-trained on proprietary data by Meta \cite{grattafiori2024llama3herdmodels}. In contrast, our initial BitNet b1.58 8B checkpoint, based on LLaMA 3 8B, was further fine-tuned on an additional 100B tokens from FineWeb-edu \cite{penedo2024finewebdatasetsdecantingweb}. Subsequently, the Hedgehog and Hedgehog with sliding window models were trained on 200M tokens \cite{zhang_hedgehog_2024, zhang_2024_lolcats}. Since extremely low-precision weights currently need significant additional pretraining (approximately 8,700 GPU hours for BitNet b1.58 8B) across all model weights, we had no choice but to begin from the available BitNet b1.58 8B checkpoint before attention distillation, creating what is likely a ``quality of data moat'' between our resulting ELiTeFormer checkpoint and the baseline LLaMA 3 models. 


\subsection{ELiTeFormer Block-wise FPGA Simulations}
\label{sec:simulation}

We test our proposed accelerator platform for ELiTeFormer via HLS in a few environments. First, we examine the characteristics of ELiTeFormer blocks when generating register transfer-level (RTL) code and simulating via the open-source Bambu HLS 2021 \cite{ferrandi_2021_bambu} and Verilator, targeting the timings of the programmable logic on the Xilinx VCK5000 board\footnote{The VCK5000 board hosts a XCVC1902 Versal device that includes FPGA programmable logic (PL) and the adaptable intelligent engines. We only use the reconfigurable logic in our experiments}. Second, we present simulation results for ELiTeFormer, utilizing the same version of Vitis HLS (version 2021.2) \cite{amd_2025_vitishls} for RTL generation, and targeting the same board. Third, we examine the performance of ELiTeFormer when deployed to an actual VCK5000, employing Xilinx's Vivado flow for placement, routing, and deployment to the board. This section will cover the block-wise simulations, functioning as a limited ablation study.

To acquire a more granular understanding of how individual ELiTeFormer blocks perform, it is worthwhile to execute iterative simulations. To this end, we build an implementation in C, augmented with limited pragmas and directives supported by Bambu HLS 2021 \cite{ferrandi_2021_bambu}. We test two blocks that provide insight into the efficiency capabilities of ELiTeFormer. First, we synthesize and test the BitNet b1.58-styled linear projections that ELiTeFormer features against a typical full-precision linear projection. 
Second, we compare the attention blocks of ELiTeFormer and Llama 3 against one another, measuring the efficiency gains in throughput and cache compression of ELiTeFormer's efficient attention block.

Results are provided for both simulations in Figure~\ref{fig:simulation_results}, where ELiTeFormer's BitNet b1.58-styled projections perform exceptionally well in terms of accelerating the block, on average achieving a $9.6\times$ in clock cycles. Similar results are observed for accelerating $attn$ (without the output linear projection, only core attention calculations), where the efficient attention block of ELiTeFormer is very competitive at longer context lengths. This results in up to a $4.4\times$ speedup in cycles compared to a typical LLaMA 3 attention block, while also compressing the KV cache by about $12.8\times$. Collectively, these simulations confirm the high efficiency potential of ELiTeFormer in custom hardware.

\subsection{Simulating and Deploying ELiTeFormer on FPGAs}
\label{sec:deployment}
Beyond just designing FPGA implementations of specific blocks, we also simulate and deploy a platform to accelerate ELiTeFormer as a whole, targeting a VCK5000 to test its efficacy. To this end, we compare our accelerator architecture to runtime observations on a LLaMA 3 model running on an NVIDIA A100 GPU and an optimized BitNet b1.58 model running on an A100 GPU and an Intel Xeon Platinum 8168 CPU with 20 cores. In latency and throughput-related experiments, we controlled for overhead and latency common to all platforms (such as data movement) but did not control for platform-specific latency. This applies mostly to GPU baselines, which can include kernel launch overhead and any just-in-time (JIT) compilation overhead in their latency.

Our ELiTeFormer accelerator is written in C and employs Vitis HLS to generate RTL before being fed into the Vivado toolchain for synthesis and FPGA deployment. Evaluation results are gathered via the Xilinx Runtime library \texttt{xrt} for the VCK5000 and typical profiling in Python for the A100. Where appropriate, we employ both naive and optimized inference frameworks including vLLM \cite{kwon2023efficient} for optimized LLaMA 3 baselines and \textit{bitnet.cpp}\footnote{https://github.com/microsoft/BitNet} for BitNet b1.58 baselines.
Our simulations are at 8B parameter scales, but our deployed implementation scales down model size ($d_{FFN} = 8096, L = 6)$ to better reflect the capabilities of the VCK5000, which only features DDR memory, and emulates a low-latency serving scenario (we also scale down models profiled on the A100 for an apples-to-apples comparison). 
This results in an end-model size of around 1B parameters and a memory footprint of 0.19 GB from the model weights.

\subsubsection{ELiTeFormer 8B FPGA Simulation Results}
We provide some brief simulation results regarding latency and throughput for the ELiTeFormer accelerator at the 8B parameter scale, at $2^{14}$ tokens of context, and with a $mB$ of 2, targeting a VCK5000 board with programmable logic at 275 MHz, based on timing closure estimations from Vitis HLS. Table~\ref{tab:8b_sim} showcases the comparison between a simulated ELiTeFormer 8B to LLaMA 3 8B running on an NVIDIA A100 and using vLLM for inference, as well as a fairly optimized BitNet b1.58 8B checkpoint using \texttt{bitnet.cpp} as a backend. Not only does the simulated ELiTeFormer accelerator improve on LLaMA 3 run via vLLM latency by $4.5\times$ at long contexts, but it also improves on the baseline's throughput by $2.2\times$. Improvements upon BitNet b1.58 with a \texttt{bitnet.cpp} backend are more modest, but we still achieve a latency improvement of $2.7\times$. The throughput of the simulated platform is worse than that of BitNet b1.58. However, when normalizing performance by scaling the memory-bound design from the VCK5000's DDR bandwidth to the Xilinx U280's HBM bandwidth, we anticipate seeing ELiTeFormer achieve a $5.8\times$ improvement in throughput over BitNet b1.58. These results speak further to the potential performance of ELiTeFormer at larger scales.

\begin{table}[t]
    \centering
    \caption{Brief latency and throughput results based on the ELiTeFormer 8B accelerator simulations derived from system estimations from Vitis HLS targeting a Xilinx VCK5000. Latency is based on a single forward pass during generation for a single sample batch, isolating for latency related to data movement but not for GPU-specific overhead. Throughput is based on the average of 5 tokens for a batch size of 32. Normalized throughput is provided by scaling the bandwidth of the simulated device for ELiTeFormer to Xilinx U280 HBM bandwidth (the VCK5000 is DDR only). All measurements are for a context length of $2^{14}$. Best values are bolded.}
    \begin{tabular}{l|ccc}
        \multirow{2}{*}{Arch. and Device} & Latency & Thrpt. & Norm. Thrpt.\\
        & (ms) & (tok/s) & HBM (tok/s) \\
        \hline
        Llama 3 8B, A100 (vLLM) & 544 & \hphantom{0}5.71 & \hphantom{0}5.71 \\
        BitNet b1.58 8B, A100 (bitnet.cpp) & 319 & \textbf{14.12} & 14.12\\
        ELiTeFormer 8B (Ours) & \textbf{120} & 12.52 & \textbf{82.27} \\
    \end{tabular}
    \label{tab:8b_sim}
\end{table}

\begin{figure}[t]
\begin{tikzpicture}[scale=1.0]
\begin{loglogaxis}
[
    xlabel={Context Length (tokens)}, 
    ylabel={Latency (ms)}, 
    legend pos=north west,
    grid=major,
    tick align=outside,
    ytick pos=left,
    ytick={1e2,1e3},
    yticklabels={$10^2$, $10^3$},
    extra y ticks={2000,3000,4000, 5000},
    extra y tick labels={,,,},
    extra y tick style={
        grid=none,
        major tick length=3pt,
    },
    xtick pos=bottom,
    log basis x=2,
    ymax=5000,
    ymin=100,
    legend cell align={left},
    legend style={nodes={scale=0.80, transform shape}},
]
     \addplot coordinates{
        (2048, 375)
        (4096, 382)
        (8192, 440)
        (16384, 505)
    };
    \addplot coordinates{
        (2048, 115)
        (4096, 147)
        (8192, 188)
        (16384, 291)
    };
    \addplot coordinates{
        (2048, 255)
        (4096, 517)
        (8192, 1000)
        (16834, 3686)
    };
    \addplot coordinates{
        (2048, 150)
        (4096, 150)
        (8192, 153)
        (16384, 170)
    };
    \addplot coordinates{
        (2048, 129)
        (4096, 129)
        (8192, 129)
        (16384, 129)
    };
    \legend{Llama 3 1B on A100 (Naive), Llama 3 1B on A100 (vLLM), BitNet b1.58 1B on CPU  (bitnet.cpp), BitNet b1.58 1B on A100 (bitnet.cpp), ELiTeFormer 1B on VCK5000 (Ours)}
\end{loglogaxis}
\end{tikzpicture}
\caption{Observed latency (ms) (lower is better) on a single forward pass during generation for our ELiTeFormer 1B accelerator deployed to a VCK5000 and LLaMa 3 1B running on an A100 for varying context lengths. Latency is post initialization and isolates for data movement, but does include GPU-specific overhead. Results are the average of five runs.}
\label{fig:actual_latency}
\end{figure}
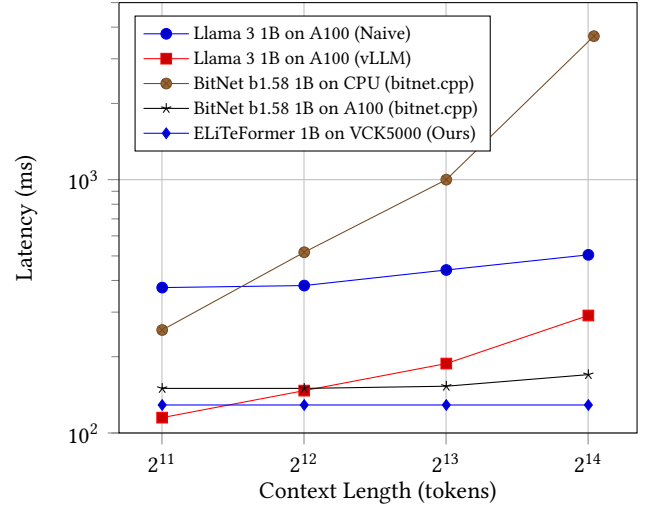


\subsubsection{ELiTeFormer 1B Latency Profile}
As previously discussed, we provide deployment results for our proposed platform with a scaled-down configuration,  which better reflects the capabilities of the VCK5000 and emulates a low-latency serving scenario.

Regarding observed latency, Figure~\ref{fig:actual_latency} shows that our proposed, accelerated solution, running at a clock frequency of 275 MHz, achieves a $3.9\times$ speedup in terms of latency at long context lengths compared to a basic LLaMA 3 implementation executing on an A100. Even when employing a more optimized implementation in vLLM \cite{kwon2023efficient}, a remarkably efficient, general framework for inference with popular LLMs, we achieve a $2.6\times$ improvement in latency. Indeed, while LLaMA 3 executed via vLLM on an A100 improves significantly on the original model baseline at shorter context lengths, the gap closes at longer context lengths. Regarding the BitNet b1.58 comparisons, the CPU implementation is competitive with a naive LLaMa 3 implementation at shorter contexts, but is not competitive with more optimized frameworks. On the other hand, the highly optimized GPU implementation is very competitive, despite still operating at a higher precision (i.e., ternary precision is not fully utilized in general-purpose hardware).. Even so, we still achieve a $1.3\times$ improvement in latency over the optimized BitNet b1.58 GPU baseline at long contexts. 

\subsubsection{ELiTeFormer 1B Energy Efficiency}
\begin{table}[t]
    \centering
    \caption{Energy efficiency results for a single forward pass during generation for our ELiTeFormer deployed to a VCK5000 and LLaMA 3 running on an A100 at $2^{14}$ tokens of context. Throughput in tokens/sec is divided by watts to form tokens/J (higher is better), serving as a composite efficiency metric. Best values are bolded.}
    \begin{tabular}{l|cc}
        Arch. and Device & Power (W) & Tokens/J \\
        \hline
        Llama 3 1B on A100 (vLLM) & 83 &  0.03 \\
        BitNet b1.58 1B on A100 (bitnet.cpp) & 80 & 0.07 \\
        ELiTeFormer 1B on VCK5000 (Ours) & \textbf{26} & \textbf{0.30} \\
    \end{tabular}
    \label{tab:energy_eff}
\end{table}

\begin{table}
    \centering
    \caption{Utilization values (in \%) for accelerated ELiTeFormer on a VCK5000 for the whole design relative to the platform (left) and then LUT and BRAM utilization broken down by stage relative to the design (right, stages defined in Section~\ref{sec:pipeline}). This is based off of Vitis HLS system estimations.}
    \begin{tabular}{l|c}
        VCK5000 & Resource \\
        Resource & Utilization \\
        \hline
        BRAM & 41.5 \\
        DSP & 19.5 \\
        FF & 30.2 \\
        LUT & 44.7 \\
    \end{tabular}
    \qquad
    \begin{tabular}{l|cc}
        ELiTeFormer  & LUT & BRAM \\
        Stage & Utiliz. & Utiliz. \\
        \hline
        $QKV$ & \hphantom{0}5.4 & 17.0 \\
        $attn$ & 60.5 & 80.4 \\
        $FFN_{1, 2}$ & 22.0 & \hphantom{0}3.1 \\
        $FFN_3$ & 11.2 & \hphantom{0}1.5 \\
    \end{tabular}   
    \label{tab:util}
\end{table}

\begin{table*}
    \centering
    \caption{High-level comparison of various works that propose LLM inference acceleration in FPGAs. Headings marked with an asterisk (*) indicate that some adaptation may be necessary for the accelerator to be applicable, such as varying degrees of fine-tuning, pruning, and sparsification. For numerical values, higher is better. 
    Normalized throughput is calculated based on scaling these accelerators assuming they are memory bound (unless otherwise stated in the work) to U280 HBM bandwidth (if the card was not used) and to 8B parameters in a proportional manner. 
    }
    \begin{tabular}{l|cccc|ccc}
        Accelerator Platform & \multirow{1}{*}{Accelerates} & \multirow{1}{*}{Accelerates} & \multirow{1}{*}{Accelerates} & \multirow{1}{*}{Specialized for} & \multirow{1}{*}{Model} & \multirow{1}{*}{Cache} & \multirow{1}{*}{Normalized}  \\
        & Full LLM Arch. & Generic LLM* & Ternary LLM* & Long Context &  Compr. Ratio & Compr. Ratio  & Thrpt. (tok/s) \\
        \hline
        TransFRU\footnotemark[6] \cite{wang_transfru_2024} & \xmark & \cmark & \xmark & \xmark & \hphantom{0}4.0$\times$ & \hphantom{0}4.0$\times$ & \hphantom{0}4.83 \\
        FlightLLM \cite{zeng_flightllm_2024} & \cmark & \cmark & \xmark & \xmark & \hphantom{0}4.6$\times$ & \hphantom{0}2.0$\times$ & 48.12 \\
        ELiTeFormer (Ours) & \cmark & \xmark & \cmark & \cmark & 10.1$\times$ & 12.8$\times$ & 82.27 \\
    \end{tabular}
    \label{tab:platform_comparisons}
\end{table*}

We compare average energy efficiency for both runtime environments in Table~\ref{tab:energy_eff}, which shows a $3.2\times$ improvement in average power consumption for our ELiTeFormer accelerator on a VCK5000 compared to LLaMA 3 on an A100. This data was gathered by profiling during runtime with \texttt{xbutil} and \texttt{nvidia-smi} for VCK5000 and GPU runs (which report average power consumption during an execution), respectively. 
Moreover, we also include a composite metric for energy efficiency (ignoring device area) in tokens/J, where throughput is defined as tokens/sec for the models being tested, and we measure the device average power while running those layers. As observed in Table~\ref{tab:energy_eff}, our ELiTeFormer accelerator achieves a $5.6\times$ gain in the proposed composite metric, underscoring just how drastically more efficient it is than the LLaMA 3 on an A100 baseline. 

\subsubsection{ELiTeFormer 1B Resource Utilization Breakdown}
Resource utilization results are presented in Table~\ref{tab:util}, demonstrating that the implemented design efficiently utilizes resources.
The utilization of LUTs and BRAMs is also broken down by stage in Table~\ref{tab:util}, which illustrates differing per-stage resource consumption profiles. The $attn$ stage is particularly resource-intensive compared to the other stages because it is responsible for a series of complex floating-point computations, even if the computations themselves are relatively low-latency. The $FFN_{1,2}$ stage exhibits approximately double the resource utilization of the $FFN_3$ stage, which is expected, as $FFN_{1,2}$ accounts for two parallel linear projections, versus only a single linear projection in $FFN_3$. 

\section{Related Work}
There is a range of prior solutions for customized Transformer acceleration. We review important related work in this section, outlining their core contributions and advantages.

\subsection{Comparisons to Recent FPGA Transformer Accelerators}
\label{sec:comparisons}

Two contemporary works stand out as competitive regarding accelerating LLM-related workloads via FPGAs: TransFRU \cite{wang_transfru_2024} and FlightLLM \cite{zeng_flightllm_2024}.

TransFRU focuses on accelerating only the classical attention operator, leveraging mixed precision and attention activation sparsity via sorting. It rightfully notes that attention is a bottleneck in long-context workloads for models like BERT \cite{devlin2019bertpretrainingdeepbidirectional} and around its size. Their proposed acceleration is possibly synergistic with the sliding window component of our attention mechanism, but we would also emphasize that their analysis of attention being a bottleneck occurs more often for the smaller models they tested, especially on bidirectional models like BERT. We deploy at larger scales, only on generative workloads, and accelerate attention first from an algorithmic perspective with hybrid attention.

On the other hand, FlightLLM is the current state-of-the-art for FPGA-based LLM acceleration. Unlike this work, they propose accelerating both prefilling and generation with a unified architecture. In comparison, we propose leveraging a heterogeneous inference environment where another device performs resource-intensive prefilling and our FPGA-based accelerator performs generation, providing a better overall fit for FPGAs as a device. That being said, FlightLLM does achieve impressive results, although there is some overhead to accommodate their custom ISA (around 3.25 GB of DRAM). When deployed on a Xilinx Alveo U280, they observe a peak throughput of 55 tok/sec for multi-batch workloads at low context lengths.

We compare our proposed ELiTeFormer and accelerator architecture to these works in Table~\ref{tab:platform_comparisons}, using our 8B parameter simulation results for throughput and proportionally normalizing to U280 bandwidth for our memory-bound design. While the platforms above benefit from being applicable to more general LLM architectures, they are unable to properly accelerate ternary architectures, do not specialize in long-context modeling and acceleration, and exhibit subpar model and KV cache compression ratios. Moreover, when normalizing the performance of these platforms in accordance with U280 HBM bandwidth (we use ELiTeFormer 8B's high-throughput simulation results for this data point), we demonstrate superior throughput as well. 
As a brief note, throughput comparisons here do not normalize for differing context lengths because prior works often report results at short contexts or omit the context length entirely. This omission is unfavorable to ELiTeFormer, which achieves its strongest performance at long contexts. We also do not account for differences in resource usage (e.g., DSP consumption), a factor that would advantage ELiTeFormer because its ternary blocks require no DSPs.

\footnotetext[6]{TransFRU does not have its throughput scaled by parameter count due to only accelerating attention. Instead, it is scaled by embedding dimension and layer count.}

\subsection{Other Transformer Acceleration Efforts}

While specific accelerators were previously discussed in Section~\ref{sec:comparisons}, we cover this topic more generally here. 
When it comes to accelerating Transformers on FPGAs, some existing work has explored challenges and opportunities in this area. For example, some works have investigated the potential for FPGAs to exploit complex pruning behavior \cite{peng_accelerating_2021, shi_fitop_2024, wang_efficient_2024}. Other works have investigated holistic optimizations, most of which were only rendered possible via fine-grained optimizations provided in FPGAs \cite{wang_via_2022, sarker_edge_moe_2023}. Similar to investigations rooted in ASIC designs, several works have explored leveraging more dynamic sparsity in FPGAs for attention acceleration \cite{chen_dynamic_nm_2023, fang_efficient_2022, fang_nm_2022}. Some prior work has examined the viability of constructing Transformer-based accelerators via HLS \cite{plagwitz_trac_2022}. Finally, one prior work did attempt to accelerate linear attention in an efficient ASIC design \cite{dass_vitality_2023}.


\subsection{Efficient Attention Alternatives}
A range of alternative attention operators focused on computational efficiency have been proposed to replace attention. 
Sparsity-focused attention alternatives were among the first to be proposed \cite{child_sparsetransformer_2019, bigbird2020, beltagy_longformer_2020}, focusing on ignoring certain attention activations to improve computational efficiency.  
Other common sub-quadratic options tend to feature low-rank approximations or bucketizing in some fashion \cite{wang_linformer_2020, kitaev_reformer_2020, peng_abc_2022}. Linear attention, as opposed to simply sub-quadratic options, generally replaces the softmax operator in typical attention with separable kernel functions such that computation can be reordered into an RNN-like hidden state-based structure, resulting in linear runtime and memory complexity \cite{kathrapalous_linear_2020, agostinelli_improving_2023, yang_gla_2024, agostinelli_leapformer_2024, zhang_hedgehog_2024, xiong_nystromformer_2021, performer2020, yang_delta_2024, kacham_polysketch_2024}.

\section{Conclusion}
In this work, we introduce ELiTeFormer, a pioneering efficient Transformer-based LLM model that integrates hybrid linear attention with ultra-low-precision linear projections, achieving unprecedented model weight and key-value cache compression, making it suitable for LLMs deployed in constrained or large-batch environments.
ELiTeFormer achieves only 3.0\% MMLU degradation compared to BitNet b1.58 while maintaining $10\times$ model and achieving $12.8\times$ cache compression.
We propose a novel PE for low-precision linear projection computations that efficiently exploits the ternary weights of our model.
Additionally, we synthesize, simulate, and deploy a co-designed, highly customized accelerator for ELiTeFormer on a Xilinx VCK5000. We observe, on average, a $9.6\times$ speedup in our FFN blocks and a $4.4\times$ speedup in our attention blocks compared to LLaMA 3. When deploying our proposed accelerator, we achieve a $3.9\times$ speedup in terms of latency and a $3.2\times$ improvement in energy efficiency compared to LLaMA 3 running on an NVIDIA A100 GPU.
These accomplishments underscore the potential of reconfigurable hardware and ELiTeFormer for accelerated, efficient LLMs.

\begin{acks}
This work was supported by the US Department of Energy, Office of Science, Office of Advanced Scientific Computing Research's Computer Science Competitive Portfolios program and the Advanced Memory to support Artificial Intelligence for Science (AMAIS) at Pacific Northwest National Laboratory (PNNL).
\end{acks}



\bibliographystyle{ACM-Reference-Format}
\bibliography{refs}

\end{document}